\documentclass{sig-alternate}

\usepackage{booktabs} 
\usepackage{enumitem}
\usepackage{amsmath}
\usepackage{algorithm}
\usepackage[noend]{algpseudocode}
\usepackage{algpseudocode}
\usepackage{booktabs} 
\usepackage{soul}
\usepackage{color}
\usepackage{times}
\usepackage{graphicx}
\usepackage{url}
\usepackage{xcolor}
\usepackage{stfloats}
\usepackage{caption}
\usepackage{authblk}

\title{Discovering Latent Patterns of Urban Cultural Interactions in WeChat for Modern City Planning}

\newcommand{\affiliationfont}{\fontsize{8}{8}\selectfont}

\author[1]{Xiao Zhou}
\author[2]{Anastasios Noulas}
\author[1]{Cecilia Mascoloo}
\author[3]{Zhongxiang Zhao}
\affil[1]{\affiliationfont Computer Laboratory, University of Cambridge, UK}
\affil[2]{\affiliationfont Center for Data Science, New York University, New York, NY, USA}
\affil[3]{\affiliationfont WeChat Business Group, Tencent Inc. , Beijing, China}

\begin{document}
\maketitle

\begin{abstract}
Cultural activity is an inherent aspect of urban life and the success of a modern city is largely determined by its capacity to offer generous cultural entertainment to its citizens. To this end, the optimal allocation of cultural establishments and related resources across urban regions becomes of vital importance, as it can reduce financial costs in terms of planning and improve quality of life in the city, more generally. In this paper, we make use of a large longitudinal dataset of user location check-ins from the online social network WeChat to develop a data-driven framework for cultural planning in the city of Beijing. We exploit rich spatio-temporal representations on user activity at cultural venues and use a novel extended version of the traditional latent Dirichlet allocation model that incorporates temporal information to identify latent patterns of urban cultural interactions. Using the characteristic typologies of mobile user cultural activities emitted by the model, we determine the levels of demand for different types of cultural resources across urban areas. We then compare those with the corresponding levels of supply as driven by the presence and spatial reach of cultural venues in local areas to obtain high resolution maps that indicate urban regions with lack of cultural resources, and thus give suggestions for further urban cultural planning and investment optimisation.
\end{abstract}


\section{Introduction}

The opportunity to enjoy cultural and entertainment activities is an essential element of urban life. Cultural spending is a considerable fraction of the annual budget of a city, as it is a key catalyst of social life and an important quality of life indicator in urban environments. Typically, in megacities like Beijing, London or New York, the financial resources allocated to support cultural events and related urban development (e.g. museum construction and maintenance) is in the order of tens of millions of dollars annually~\cite{London}. 
Furthermore, in today's inter-connected world, the opportunity to experience a diverse set of cultural activities is amplified; citizens equipped
with mobile devices can utilise a wide range of mobile applications and services that inform them on on-going social and cultural events, as well as on the best areas in the city to explore culture. In this setting, urban culture explorers are also mobile users who emit digital traces of where and when they are participating in cultural events.
The new window onto the cultural life of a city, opened by the availability of novel sources of mobile user data, paves the way to the development of new monitoring technologies of urban cultural life. This can power evidence-based cultural policy design and optimise urban planning decisions. As an example, by tracking the interactions of mobile users with cultural venues across space and time, fine-grained indicators of areas in the city where there is lack, or excessive supply, of cultural resources can be devised. 

In this paper, we exploit a large set of spatio-temporal footprints of users in the online social network WeChat to obtain patterns of urban cultural interactions in the city of Beijing. We first devise a latent Dirichlet allocation (LDA)~\cite{blei2003latent} based method that takes as input check-ins at venues over time to identify clusters of mobile users with common cultural profile patterns. Having identified the geographic spread of check-in activity of each user cluster, we then estimate the primary locations of users in a cluster, in terms of home or work, and evaluate the degree of accessibility to cultural venue resources for a user. Next, we empirically demonstrate how the supply-demand balance of cultural services in a city can be highly skewed and we pinpoint in cartographic terms the areas in the urban territory where supply could improve through appropriate investment. In more detail, we make the following contributions: 
\begin{itemize}[leftmargin=0.3cm]
	\item \textbf{Cultural patterns extraction from check-in data:}
	We obtain raw representations of time-stamped check-in data at WeChat venues and use those as input to a novel extended version of the standard LDA model that takes into account temporal information (TLDA) on when venues of certain cultural categories are visited. We evaluate the performance of the model with a novel metric of coherence between top cultural venue categories observed in a cluster of users and the time periods of activity that are characteristic to each pattern. Using the metric we optimise the TLDA model and identify the presence of six latent cultural patterns in Beijing, each of which bears characteristic spatio-temporal manifestations of user activity at cultural venues. The description of the TLDA model and the corresponding data representations are described in Section~\ref{sec:patterns}.
	
	\item \textbf{Determining cultural demand patterns of users spatially:}
	Having obtained the latent cultural patterns through the TLDA model, we then exploit the frequencies of user check-ins across space to determine the \textit{levels of demand} of cultural activities in different areas of the city. In this context, we present POPTICS, a user-personalised version of the OPTICS algorithm~\cite{ankerst1999optics}, used here to identify clusters of user activity hotspots. These primary locations of user activity are the means to quantifying demand levels for cultural resources spatially. Overall, the output of this process corresponds to a set of heat maps depicting the intensity levels of user activity geographically for each of the cultural pattern emitted by the TLDA model. We consider such intensity levels to reflect user driven demand of cultural resources geographically and present our results in Section~\ref{sec:demand}.
	
	\item \textbf{Identification of areas that lack cultural offering:}
	In addition to obtaining spatial descriptions of the demand levels for each cultural pattern observed in the city, we determine the \textit{supply levels} of cultural resources using the spatial distribution of cultural venues and users' check-ins belonging to each TLDA pattern as input. For each region in the city, we obtain a demand-supply ratio (DSR), high values of which are indicative of an area lacking cultural venues, whereas low values suggest oversupply of cultural establishments in a region. In Section~\ref{sec:experiments} we generate precision maps of such supply and demand patterns for each cultural pattern and demonstrate how users who live in high-DSR neighbourhoods but adhere to a specific cultural pattern travel longer distances in the city to access the resources they are interested in. The latter is an indication of how lack of resources in an area translates to larger travel distance for its residents.
\end{itemize}

In summary, the methodology put forward in the present work paves the way to novel data driven urban cultural planning schemes that dynamically adapt to the profiles of residents  active in city regions. Such schemes exploit the rich characteristics of new digital data sources and have the potential to improve over planning decisions that currently tend to rely solely on residential population density and are agnostic to both user preferences and fine grained temporal signatures of user behaviour.

\section{Related Work}
The flourishing of location technology services (LTS)~\cite{hasan2014urban,bauer2012talking} and the distinct advantage of topic modeling~\cite{blei2003latent} in uncovering latent patterns have encouraged researchers to employ them as data sources and method, respectively, in large-scale urban human mobility studies. From the perspective of the users, \cite{jiang2015author} proposed a topic-based model to recommend personalized points of interest (POIs) for tourists. \cite{yuan2013we} leveraged topic modeling to learn lifestyles for individuals based on their digital footprints and social links. \cite{kurashima2013geo} established a geo-topic model for location recommendation with the consideration of activity area of users. \cite{yin2013lcars} proposed a recommender system for both venues and events according to personal preferences. From urban computing perspective, \cite{bauer2012talking} analysed the comments published alongside Foursquare check-ins to detect the topics in neighbourhoods. \cite{yuan2012discovering} utilised human mobility flows between POIs to identify land use functions for urban regions. 

Even though topic modeling has been shown as an effective unsupervised approach to discovering latent mobility patterns in existing literature, it still has some limitations. Firstly, urban activities are time-sensitive that citizens show strong time preferences for different activities  \cite{hasan2014urban,bauer2012talking}. However, these temporal behavioural patterns in urban daily life can hardly be captured by topic models as classical topic modeling does not have time factors integrated\cite{chen2017effective}. Some existing works considering temporal characteristics did not modify the model structure fundamentally~\cite{bauer2012talking}, leaving to the interactions between urban activities and temporal features largely untapped. Secondly, evaluation methods are missing in the application of topic modeling to human mobility study. The number of topics was either selected without testing \cite{hasan2015location,bauer2012talking,yuan2013we} or by using traditional evaluation methods in text mining cases directly \cite{hu2013spatio}. Thirdly, few published research findings obtained through topic modeling can be used in practical urban planning.  And finally, current publications focus on mining the patterns of all types of urban activities in general, without looking at particular subgroups of people or urban venues.  

Our research is different from existing works in the following aspects: i) We integrate time parameter with standard topic model to gain insights into temporal behavioural patterns along with cultural activities of people. ii) We device a novel evaluation method for the temporal topic model to value its performance quantitatively. iii) We detect the heterogeneity on various cultural demand and supply levels in the city environment. To the best of our knowledge, this is the first methodological work for urban cultural patterns mining using large-scale location services data in practical urban cultural planning. 

\begin{figure*}
    \centering
	\includegraphics{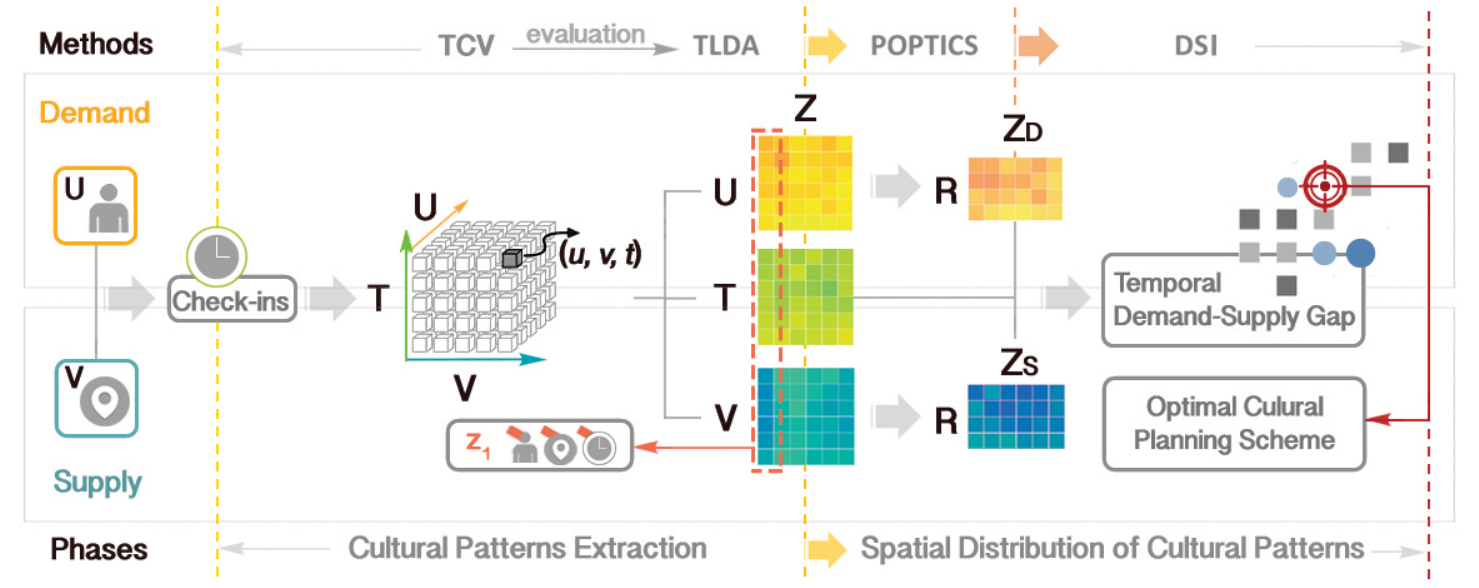}
	\caption{Description of Main Phases in the Research.}
	\label{fig:framework}
\end{figure*}

\section{Approach at a Glance}

The research framework outlined in Figure~\ref{fig:framework} includes two main phases: cultural patterns extraction, followed by spatial distribution of cultural patterns. These two phases as a whole provide an integrated approach to optimising the cultural resources allocation in cities and refining the urban cultural planning scheme by using LTS data as input. To this aim, we propose three models tailored to the characteristics of spatio-temporal social network data.

\noindent{\bf Cultural Patterns Extraction}:
Cultural patterns extraction begins with raw check-in data processing. We build the tuple $ (u,v,t) $ to denote each qualified check-in record, indicating user $ u $ visited venue category $ v $ at time $ t $. The whole set of check-ins formatted in this way are stored in a data cube $(U,V,T)$ displaying the check-in history of users in the city before being sent to the TLDA model as input data. At this stage, temporal coherence value (TCV) measurement is designed to evaluate the performance of TLDA and choose the optimal input parameter $ K $, which denotes the number of latent patterns $ Z $. The TLDA model applied in this research enables us to essentially group $ U $, $ T $, and $ V $  into $ K $ cultural patterns and discover the associations between users, time, and various cultural activities.

\noindent{\bf Spatial Distribution of Cultural Patterns}:
Building on the output of TLDA, we further devise the POPTICS algorithm and the demand-supply interaction (DSI) model to get a general picture of how cultural check-ins and venues belonging to different patterns distribute spatially in the city, so as to uncover the gap between user-side demand and venue-side supply for different cultural patterns. Our primary assumption here is that cultural demand of users and supply capability of cultural venues are heterogeneous across regions that by mapping the demand-supply ratio, we can infer whether a type of cultural resources is insufficient or over-supplied relatively in the city.

\section{Datasets}\label{dataset}
WeChat\footnote{http://www.wechat.com/en} is a mobile social application launched by Tencent in January 2011 and has currently become the most popular mobile instant messaging application dominating the Chinese market. According to the latest data report published by the WeChat team at the Tencent Global Partners Conference~\cite{WeChat17}, by the end of September 2017, it had 902 million average daily logged in users sending 38 billion messages in total every day. WeChat's 'Moments' function, which is an equivalent of Facebook's timeline feature, allows users to share their status or anything of interest via photos, text, videos, or web links with their contacts. When a user posts on Moments, a time stamp will be generated automatically. Additionally, the user has the option to share their current place from a list of pre-selected locations nearby in WeChat. This real-time location-based service provided by 'Moments' depicts the routine lives of users at a fine-grained spatio-temporal scale, and provides a precious dataset that enables us to discover urban activity patterns.

\noindent
\textbf{Advantages of WeChat Dataset.} For our analysis, the WeChat dataset possesses a number of natural advantages:

\begin{itemize}[leftmargin=0.3cm]
	\item \textit{High population coverage levels in cities.} According to a survey by Tencent in September 2015, 93\% of the population in the first-tier cities in China\footnote{Beijing, Guangzhou, Shanghai and Shenzhen} were WeChat users. As for the selected case study in this research, Beijing has 21,136,081 monthly active users\footnote{users who have logged into WeChat within the month} according to our statistics, making up 97.4\% of the residents\footnote{The permanent population of Beijing is 21.907 million by the end of 2017 according to Beijing Municipal Bureau of Statistics. http://www.bjstats.gov.cn/tjsj/yjdsj/rk/2017} in September 2017. This high popularity makes the observation of users' mobility patterns through the lens of WeChat a reliable proxy to the real mobility of Beijing residents.
	\item \textit{Wide age distribution of users.} Compared with many other social media services, the WeChat penetration among middle-aged and senior users is relatively high. Although people born in the 80s and 90s are still the major groups, the monthly active WeChat users between 55-70 years old in September 2017 were approximately 50 million. In other words, WeChat is a representative social media data source reflecting a wide range of age groups in the general population. 
	\item \textit{Private circle visiblity.} With an in-group design, the social circle of a user on WeChat is mainly comprised of relatives, friends, and colleagues who have a close relationship with him in life. Additionally, WeChat empowers the user with the right to control over exactly who has access to each single post on his 'Moments'. This powerful feature creates a secure and private environment that encourages WeChat users to communicate freely and share their check-ins.
\end{itemize}

\noindent
\textbf{Moments Check-in Data.} The main dataset employed for this study comes from  the anonymized logs of complete WeChat Moments posting activities. We collected all check-in records with venue information provided in Beijing during the four months of October 2016, January, April, and July 2017. In total, there were 56,239,429 check-ins created by 9,517,175 users at 2,428,182 venues. For each check-in, information about user ID, time stamp, coordinates, and POI category is provided. Through the lens of this dataset, we are thus able to observe who visited where at what time for what purpose. We obtained IRB approval in the University of Cambridge to work on the data for the purpose of this paper.

\section{Cultural Patterns Extraction}
\label{sec:patterns}
In this section, we present how urban cultural patterns can be extracted from Moments check-in records. Before discussing the issue in a more depth, we first state the meaning of culture and culture-related terms for the purpose of this work. 

\noindent
\textbf{Cultural Venues.} Cultural venues are defined as urban places of arts, media, sports, libraries, museums, parks, play, countryside, built heritage, tourism and creative industries, following the line set by the Office of the Deputy Prime Minister in Regeneration through Culture, Sport and Tourism~\cite{zhou2017cultural, ODPM99}. Based on this definition, 37 categories of WeChat cultural venues are selected in this research.

\noindent
\textbf{Cultural Check-ins.} The check-in activities taking place at the cultural venues are called cultural check-ins.

\noindent
\textbf{Cultural Fans.} To classify individual cultural patterns, we only focus on users who have a certain minimum level of cultural check-ins during the observation time and call them cultural fans. 

In the following subsections, we illustrate the necessity of considering temporal factors in urban cultural patterns mining, before introducing the TLDA model which integrates temporal characteristics with a particular subgroup of cultural activities, followed by the introduction of a novel evaluation method for the TLDA.

\subsection{Temporal Factors Extraction}
\begin{figure*}
	\includegraphics{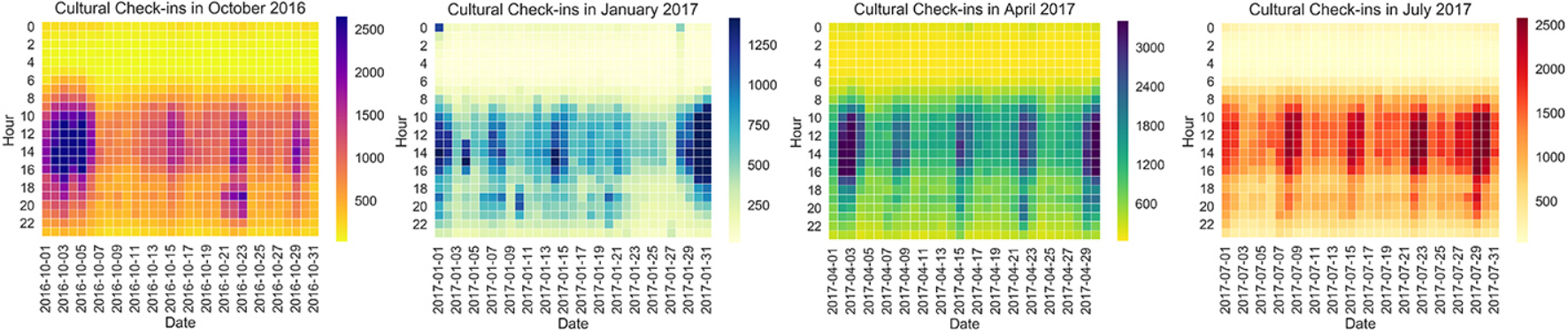}
	\caption{Temporal Cultural Check-in Distribution of Four Months in Beijing.}
	\label{fig:temporal}
\end{figure*}

In Figure ~\ref{fig:temporal} we show temporal cultural check-in distribution of the four selected months in Beijing. The purpose of creating these heat maps in a calendar format is to present the cultural check-in frequency by date and hour corresponding to four seasons chronologically. In each subfigure, the date is plotted along the horizontal axis with hour appearing on the vertical axis to unveil cultural visiting patterns associated with temporal factors, which will later be explored in further depth.

It can be seen from Figure~\ref{fig:temporal} that hourly and weekly cultural visiting patterns are both significant in general. For all the seasons, the least likely hours for cultural check-in creation is during the night, from 0 to 6am. After that, the hourly frequency of cultural check-ins increases gradually and stays at a relatively high level during the daytime. On a daily basis, two peak periods can be recognized, among which, the highest one lasts for around seven hours from 10am to 4pm while a lower peak appears between 7pm to 9pm. As for weekly patterns, we can find that the check-in frequency is significantly higher for weekends than weekdays.

It is also noticeable that hourly and weekly patterns are more evident in April and July, while less regular in October and January. This observation can be explained by comparing our calendar heat maps with the public holidays in China. At the beginning of October 2016, we can see a dramatically high cultural check-in frequency for six continuous days, when people have one week off for the National Day. It is the longest holiday after the Chinese New Year, and is also called 'Golden Week' for people to reunite with families and take trips. Then the second graph follows and presents the situation in the 'New Year' month. It can be discovered that both the first cells corresponding to the New Year's Day (01/01/2017) and the Spring Festival (28/01/2017) have distinctively higher values compared to the rest. In addition, during the week before the Chinese New Year, the number of cultural check-ins is much smaller than that in other weeks. Moreover, on 27/01/2017, the day before the Lunar New Year, the check-in frequency is particularly low, forming a sharp contrast with the following Spring Festival week. This is because typically, Chinese people prefer to stay at home with families before the New Year's Eve, waiting for the coming new year, but would like to hang out with friends in the next few days. 

\subsection{Temporal Latent Dirichlet Allocation}
\begin{figure}
	\includegraphics{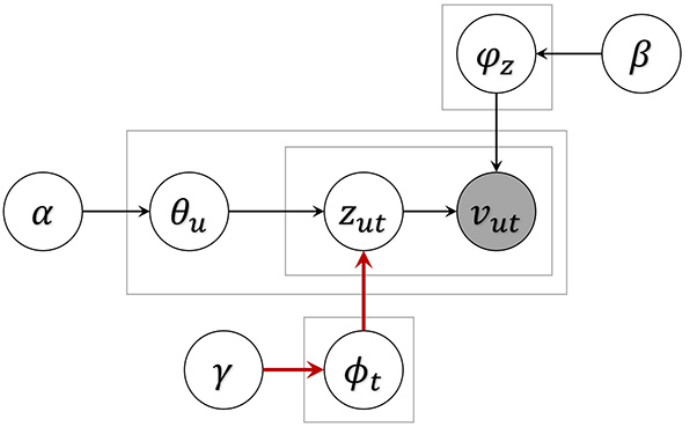}
	\caption{Graphical Model for TLDA.}
	\label{fig:TLDA}
\end{figure}

The classical LDA~\cite{blei2003latent} is a hierarchical Bayesian model which has been shown as an effective unsupervised learning method in discovering structural daily routines~\cite{huynh2008discovery,hasan2014urban,farrahi2011discovering, sun2017discovering}. However, the original LDA approach is built based on the 'bag-of-words' assumption~\cite{hasan2014urban}, which means that it only considers the number of times each word appears in a document, without involving any temporal consideration~\cite{chen2017effective}. According to the observations made in the previous subsection, the periodicity of cultural check-ins in different levels of temporal granularity is so obvious that we chose to explicitly incorporate it to the LDA model. We thus propose temporal latent Dirichlet allocation (TLDA) based on~\cite{chen2017effective} in this paper. The TLDA is an extended version of LDA that integrates time factors into the original model, so as to uncover multiple associations between users, urban activities, and their corresponding temporal characteristics. The graphical model representation of TLDA is shown in Figure~\ref{fig:TLDA}, where the lower part integrated by red arrows is the addition of TLDA. In the figure, circles represent parameters and the meaning of which are described in Table~\ref{tab:notation}.

\begin{table}
	\caption{Notation and Description}
	\label{tab:notation}
	\begin{tabular}{cl}
		\toprule
		Symbol&Description\\
		\midrule
		$ \alpha $& Dirichlet prior over the pattern-user distributions\\
		$\beta$ & Dirichlet prior over the venue-pattern distributions\\
		$\gamma$ &Dirichlet prior over the pattern-time distributions\\
		$\theta _{u}$ & pattern distribution of user $ u $\\
		$ \varphi _{z} $ & venue distribution of pattern $ z $\\
		$ \phi_{t} $ & pattern distribution of time $ t $\\
		$ z_{ut} $ & pattern of venue category of user $ u $ at time $ t $\\
		$ v_{ut} $ & venue category of user $ u $'s check-in at time $ t $\\
		\bottomrule
	\end{tabular}
\end{table}

The framework of the TLDA model consists of four hierarchical layers, including a user layer, a time layer, a venue category layer, and a cultural latent pattern layer. The cultural pattern layer is the key layer which links the other three. As its predecessor, TLDA is also a generative model, the goal of which is to find the best set of latent variables (cultural patterns) that can explain the observed data (cultural check-ins by users)~\cite{hasan2014urban}. To generate a cultural venue category, the pattern distribution of the corresponding user is sampled from a prior Dirichlet distribution parameterized by $\alpha$, $\theta _{u}\sim  Dir(\alpha$). In a similar way, the pattern distribution of time is sampled from a prior Dirichlet distribution parameterized by $\gamma$. Based on these, the pattern assignment $Z_{ut}$ of the venue category is drawn from a multinomial distribution $Z_{ut}\sim Multi(\theta _{u},\phi_{t}$).

\begin{equation}
P(z_{ut}|\alpha ,\gamma)=\sum_{\theta _{u}}P ( z_{ut}| \theta _{u})P( \theta _{u}| \alpha) \sum_{\phi_{t}}P( z_{ut}| \phi_{t} )P ( \phi_{t}| \gamma )
\end{equation}

Then, the venue category is generated by sampling $Vut\sim Multi(\varphi _z)$. $\varphi _z$ specifies the venue distribution of pattern $z$, which is drawn form a prior Dirichlet distribution parameterized by $\beta$. 

\begin{equation}
P(v_{ut}|z_{ut}) = \sum_{\varphi _z}P(v_{ut}|\varphi _z)P(\varphi _z|\beta)
\end{equation}

After that, we estimate the maximum likelihood of $v_{ut}$ of $u$ at time $t$ by summing up $\theta_{u}$, $\phi_{t}$ and $\varphi _{z}$, as shown in the following equation. 

\begin{equation}
P(v_{ut}|\alpha,\beta,\gamma)=\sum_{\theta_{u}}\sum_{\phi_{t}}\sum_{\varphi _{z}}P(v_{ut},z_{ut},\theta_u,\phi _t,\varphi _z|\alpha ,\beta ,\gamma)
\end{equation}

In the last step, we use the Gibbs sampling algorithm to estimate the probability  distributions of pattern-user, venue-pattern, and pattern-time, respectively.

\subsection{Evaluation of TLDA}

TLDA is an unsupervised learning method which requires a pre-specified number of patterns $K$. As the temporal layer is integrated into the TLDA, conventional LDA evaluation approaches cannot be used directly to the extended model. To handle this obstacle, we design temporal coherence value (TCV) to evaluate the performance of the TLDA model and to find the optimum number of cultural patterns in the city. Inspired from the coherence value (CV) measurement proposed in \cite{roder2015exploring}, the TCV introduced in this paper is used to measure the coherence level between top cultural venue categories and time periods within each pattern, before averaging them to evaluate the overall performance of the TLDA model. 

How the TCV works step-by-step is depicted in Algorithm 1. To run the model, we need three inputs: from the output of TLDA $TV$, we obtain 1) top venue categories $V^*$ and 2) top time periods $T^*$ in each cultural pattern; and 3) all the check-in activities $SW$. It should be stated that to eliminate the influence of variance between users' check-in frequencies, the input $SW$ used in TCV is constructed by a sliding window which moves over the original check-ins of all the users. In the algorithm, we firstly define a segmentation $S_{set}^{one}$ for each top venue category $v^*$ in each pattern as equation \ref{eq:4} shows. Here $S$ is used to denote the list of $S_{set}^{one}$. The total number of $S_{set}^{one}$ in $S$ is denoted by $Q$. 

\begin{equation}\label{eq:4}
S_{set}^{one}=\left \{(v^*,V^*,T^*)|v^*\in V^* \right \}
\end{equation}

For each $S_{set}^{one}$, we calculate the \textit{normalised pointwise mutual information} (NPMI) \cite{bouma2009normalized} for $v^*$-$T^*$ vector and $V^*$-$T^*$ vector, respectively. The $j$th element of the time vector $t^*_j$ and venue category $v^*$ has the NPMI:

\begin{equation}
\overrightarrow{w}(j) = NPMI(v^*,t_j^*)^{\tau}=\left ( \frac{log\frac{P(v^*,t_j^*)+\varepsilon}{P(v^*)\cdot P(t_j^*)}}{-log(P(v^*,t_j^*)+\varepsilon}\right )^{\tau } 
\end{equation}

where $P(v^*,t^*_j)$ is the probability of the co-occurrence of $v^*$ and $t^*_j$, while $P(v^*)$ and $P(t^*_j)$ mean the probabilities of $v^*$ and $t^*_j$ ($t_j^*\in T^*$), respectively. $ \varepsilon $ is added to avoid logarithm of zero, and an increase of $\tau$ gives higher NPMI values more weight. After calculating the NPMI value for each venue category according to the above formula, we aggregate them to obtain the $j$th element of the time vector of $V^*$ by the following equation.

\begin{equation}
\overrightarrow{W}(j) = \sum_{i=1}^{I}NPMI(v_i^*,t_j^*)^{\tau}
\end{equation}

where $v_i^*$ represents the $i$th venue category in $V^*$. 

Cosine similarity is then calculated between pairs of context vectors $\overrightarrow{w}_q$ and $\overrightarrow{W}_q$ to obtain the coherence score $m_q$ for each $S_{set}^{one}$ by formula \ref{eq:7}, before we average over all top venue categories in patterns to get the final TCV $\overline{m}$ for the model through equation \ref{eq:8}. 

\begin{equation}\label{eq:7}
m_q=cos(\overrightarrow{w_q},\overrightarrow{W_q})
\end{equation}

\begin{equation}\label{eq:8}
\overline{m}=\frac{\sum_{q=1}^{Q}m_q}{Q}
\end{equation}

The higher the TCV score, the better the clustering result of the TLDA model.

\makeatletter
\def\BState{\State\hskip-\ALG@thistlm}
\makeatother

\algnewcommand\algorithmicforeach{\textbf{for each}}
\algdef{S}[FOR]{ForEach}[1]{\algorithmicforeach\ #1\ \algorithmicdo}

\makeatletter
\def\BState{\State\hskip-\ALG@thistlm}
\makeatother
\begin{algorithm}
	\caption{  Temporal Coherence Value Calculation}\label{euclid}
	\begin{flushleft}
		\textbf{Input:} $VT([V_1^*,V_2^*,...,V_K^*], [T_1^*,T_2^*,...T_K^*]), SW$\\
		\textbf{Output:} $\overline{m}$
	\end{flushleft}
	\begin{algorithmic}[1]
		\State $initialize$  $S=set()$
		\ForEach {($V^*$,$T^*$) in $VT$}:
		\ForEach {$v^*$ in $V^*$}:
		\State $S_{set}^{one}=\left \{(v^*,V^*,T^*)|v^*\in V^*\right \}$
		\State $S \gets S+S_{set}^{one}$.
		\EndFor 
		\EndFor 
		\ForEach {$S_{set}^{one}$ in $S$}:
		\ForEach {$t_j^*$ in $T^*$}:
		\State $\overrightarrow{w}(j) = NPMI(v^*,t_j^*)^{\tau}$
		\State $\overrightarrow{W}(j) = \sum_{i=1}^{I}NPMI(v_i^*,t_j^*)^{\tau}$
		\EndFor
		\State $m_q=cos(\overrightarrow{w_q},\overrightarrow{W_q})$
		\EndFor
		\State \Return {$\overline{m}=\frac{\sum_{q=1}^{Q}m_q}{Q}$}
		
	\end{algorithmic}
\end{algorithm}

\section{Refined Urban Cultural Planning}
\label{sec:demand}
Different from current cultural planning frameworks, which mainly consider the population of urban areas when allocating cultural resources, we propose a refined urban cultural planning scheme based on the results of the TLDA. Our core viewpoint here is that urban regions are heterogeneous in terms of cultural demand and supply capability, which should not be treated uniformly. By employing the TLDA model, we are able to group users according to their cultural tastes, and cluster cultural venues based on their similarities derived from human mobility behaviours. Then, after aggregating the users and cultural facilities into urban regions, we can get an idea about how the cultural demand and supply are distributed spatially in the city for different cultural patterns. Moreover, through learning the supply-demand balance across regions, we can detect the areas where particular cultural services are needed, and we are thus able to provide city government with a priority list when the financial budget is compiled for culture-related planning.

\subsection{Demand Range Determination}
In this part we explore a way to determine the main activity range of individual users as reflected in their historical check-ins. More specifically, our aims are to detect valid visits for the user, map the active ranges of areas which he visits frequently, and finally, determine the centre and radius for these active ranges. Among existing clustering methods, we find OPTICS~\cite{ankerst1999optics} a suitable approach for our problem. OPTICS is an algorithm for finding meaningful density-based clusters in spatial data \cite{kriegel2011density}. This method requires two parameters as input: the maximum radius to consider, and the least number of points to form a cluster. As check-in frequencies of users can vary greatly, setting a common number of minimum points for all users is inadequate. Considering this limitation, we propose a modified version as Algorithm 2 shows, named POPTICS that defines a different threshold for each user separately.  

In POPTICS, we collect all the $N$ locations of check-ins $L_u = [l_1,l_2,\cdots,l_N]$ for each user $u$. Here places with more than one check-in records are counted repeatedly, as they are more important in the user's life and should be given higher weights. An input parameter $ \eta $ is set to denote the percentage of a user's total check-in locations $L_u$ being considered in the calculation of core distances. Here $ \eta $ is varied for different users. The core distance for location $l_i$ is defined as the Euclidean distance between $l_i$ and the $L_u \eta$-th nearest point to it, as shown in the following function.

\begin{equation}
CD(i)= min_{\eta}Dist(i,j) (j=1,2,3,...N)
\end{equation}

After calculating the core distance, for location $l_o$, we define the reachability distance from $l_o$ to $l_i$ as: 

\begin{equation}
RD(o,i)= max(CD(o),Dist(o,i))
\end{equation}

According to the reachability distances, an ordered list of locations is generated. Then, to find meaningful cluster(s) of locations and detect outliers, a threshold of maximum reachability distance, $rd_{th}$ is set according to the score derived from formula \ref{eq:11}. The lower the score, the better the chosen $rd_{th}$ is. The group of all valid points is denoted by $RD^*$ as equation \ref{eq:12} shows.

\begin{equation}\label{eq:11}
score(rd_{th})=std(RD^*)\frac{N}{len(RD^*)}
\end{equation}

\begin{equation}\label{eq:12}
RD^* = \left \{ rd_i|rd_i\in RD \quad \textrm{and} \quad rd_i< rd_{th}\right \}
\end{equation}

\begin{algorithm}
	\caption{POPTICS}\label{euclid}
	\begin{flushleft}
		\textbf{Input:} $L_u=[l_1,l_2,...,l_N], \eta $\\
		\textbf{Output:} cluster groups of locations $GL=[L_1^*,L_2^*,...,L_r^*]$\\
		\hspace*{12mm}cluster of locations $L_i^*=[l_{i1},l_{i2},...,l_{ig}]$
	\end{flushleft}
	\begin{algorithmic}[1]
		\State $\textbf{initialize}$ $CD$=$list()$, $RD$=$list(maxdis)$,$RD(0)$=$0$,\\$seeds$=${1,2,...,N}$,  $ind$=$1$, $order$=$list()$,$GL$=$list()$,$tmp\_L$=$list()$
		\For {$i=1,2,3,...,N$}:
		\State $CD \gets CD+min_{\eta}Dist(i,j)$
		\EndFor 
		\While {$seeds != \left \{\right \}$}:
		\State $seeds.move(ind)$
		\State $order \gets order+ind$
		\ForEach {$ii$ in $seeds$}:
		\State $cur\_rd \gets max(CD(ind),Dist(ind,ii)$
		\State $RD(ii) \gets min(RD(ii),cur\_rd$
		\EndFor
		\State $ind \gets  \left \{min-index(RD_{ii})|ii \in seeds\right \}$
		\EndWhile
		\State $rd_{th} \gets min\ std(RD^*)\frac{N}{len(RD^*)}$
		\For {$ii$ in order}:
		\If {$RD(ii)< rd_{th}$}:
		\State $tmp\_L \gets tmp\_L+l_{ii}$
		\Else {}:
		\State $GL \gets GL+tmp\_L$
		\State $tmp\_L.clear()$
		\EndIf
		\EndFor
		\State $GL \gets GL+tmp\_L$
	\end{algorithmic}
\end{algorithm}

\subsection{Demand-Supply Interaction Model}
In this part we display the demand-supply interaction model (DSI). Through the TLDA model, each user $u$ is labelled as a member of a particular cultural pattern $z$. Through POPTICS, the active centre $ \mu $ and radius $r$ of each user are determined. Also, the sub active range of locations belonging to pattern $z$ can be drawn, which has the same centre $ \mu $ and a smaller pattern radius $r_{u_z}$. These results allow us to link users with the urban areas, and thus give us an indication of the demand levels of different cultural types in urban regions. The assumption here is that for a certain user $u$ from a cultural group $z$, his demand for this type of cultural service is highest at the active centre $\mu$, and decays as the distance increasing until $r_{u_z}$. The attenuation pattern is depicted by a Gaussian function. For a point $x$ within user $u$'s pattern range $r_{u_z}$ in the city, the demand influence it gets from $u$ can be obtained by:

\begin{equation}
d_{u_z}(x)=Norm(x,\mu ,r_{u_z})=\frac{1}{\sqrt{2\pi r_{u_z}^2}}exp\left ( -\frac{(x-\mu )^2}{2r_{u_z}^2}\right )
\end{equation}

The total demand in terms of pattern $z$ for area $x$ is the aggregation of influences from all users in pattern $z$. 

\begin{equation}
D_z(x) = \sum_{u_{z}}d_{u_{z}}(x)
\end{equation}

Next, we turn our focus to the supply of patterns. For each venue category $v$ in pattern $z$, we calculate the supply capability of $v$ spatially. If a user $u$ once created check-in(s) at venue $v$, then the centre of the user $\mu $ is covered by the service range of venue $v$. We find all the users who had check-ins at $v$, calculate the distances between their centres with $v$. The average of the distances is set as the standard deviation for the attenuation distribution, and denoted by $\sigma$. Based on this assumption, the supply capability of cultural pattern $z$ contributed by venue $v$ in area $x$ can be obtained through:

\begin{equation}
s_{v_z}(x)=Norm(x,v_z,\sigma_{v_z})=\frac{1}{\sqrt{2\pi {\sigma_{v_z}}^2}}exp\left ( -\frac{(x-v_z)^2}{2\sigma_{v_z} ^2}\right )
\end{equation}

The total supply level of area $x$ in the city in terms of pattern $z$ can be achieved by the following equation.

\begin{equation}
S_z(x) = \sum_{v_{z}}s_{v_{z}}(x)
\end{equation}

We then define a metric called demand-supply ratio (DSR) to capture the desirability level of a certain type of cultural service $z$ in urban areas as: 

\begin{equation}
DSR_z(x)=\frac{D_z(x)}{S_z(x)}
\end{equation}

The higher the DSR is, the greater the need of particular cultural facilities, and the higher the priority of the area in the proposed urban cultural planning scheme.

\section{Experiments}
\label{sec:experiments}
Until now, we have provided a holistic framework for urban cultural studies from extracting spatio-temporal cultural patterns to refining cultural planning for the city. Next, we will employ WeChat Moments data described in Section \ref{dataset} and use Beijing as a case to present how these models can be applied jointly in practice. 

\subsection{Data Preprocessing}
We firstly filter cultural fans based on users with at least 20 check-ins at cultural venues during the observation time. After this procedure, our dataset shows that there are 1,082 cultural venues grouped in 37 categories, and 324,809 cultural check-ins created by 18,234 cultural fans in Beijing during the selected four months. Besides a venue category label, we also represent a temporal label for each cultural check-in with three levels of identifiers, including month of year, day of week, and hour of day.  Following this form of expression, a user's check-in history can be represented as (User3, ((Concert hall, JulFri20), (Golf, OctSun10), (Yoga, AprFri18)), for example. A collection of all the cultural fans' check-ins constitutes the whole corpus, which is the input data for our analysis. 

\subsection{Cultural Patterns Extraction for Beijing}

We first run the TLDA model with the optimum number of patterns $K$ given by TCV. We adopt 7 numbers from 3 to 9 as candidates, run the TLDA for 100 iterations each, and get their respective average TCV scores as shown in Figure 4. We can see that 6 gives best performance, suggesting that cultural behaviours in Beijing should be classified into six groups based on their categorical and temporal characteristics. We then select 6 as the value of $K$ and run the TLDA model again to extract cultural patterns for the city. The main output resulting from this process are three matrices: pattern-user matrix, pattern-time matrix, and venue-pattern matrix. These matrices provide us the probability distribution of users, time periods, and venue categories over 6 patterns, respectively. These outputs as a whole tell us which group of people prefer to do what type of cultural activities at what time. 

\begin{figure}
	\includegraphics{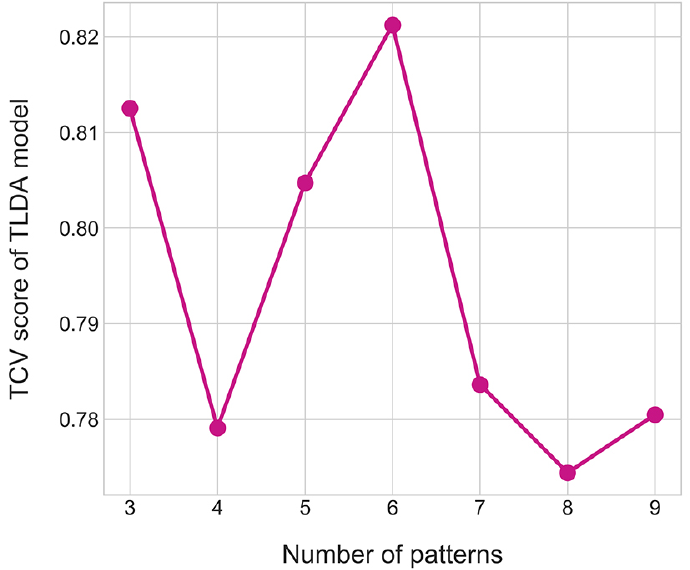}
	\caption{TCV Scores for Different Number of Patterns.}
\end{figure}

\begin{figure}
	\includegraphics{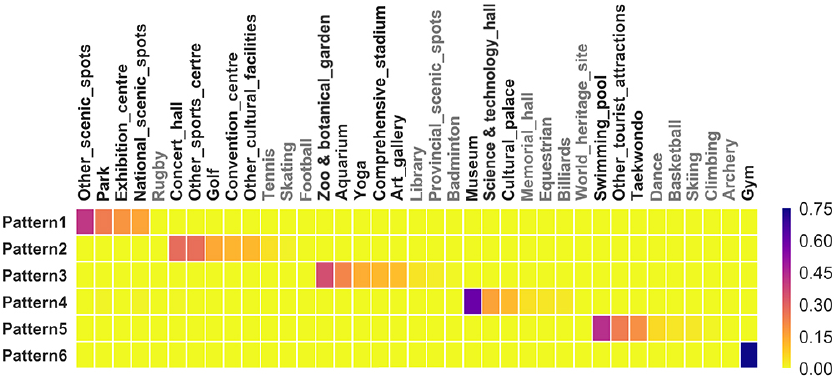}
	\caption{Probabilities of Venue Categories over Patterns.}
	\label{fig:PanVn}
\end{figure}

Figure \ref{fig:PanVn} presents the probabilities of cultural venue categories over patterns. The colour of a cell indicates the probability a category belongs to a certain cultural pattern. As we can observe, 37 cultural venue categories can be clustered separately into six cultural groups based on which pattern has the highest probability. The name of the category is printed in black as a top venue if its probability is higher than 0.1. Otherwise, it is considered not a typical category for the pattern and is coloured in grey. We also compute the cosine similarity between each pair of venue categories and show the results in Figure 6. As it can be seen in the figure, the clustering result of venues is desirable in the sense that all the within-group similarities are higher than 0.9, while most of the inter-group similarities are lower than 0.1. 

\begin{figure}
	\includegraphics{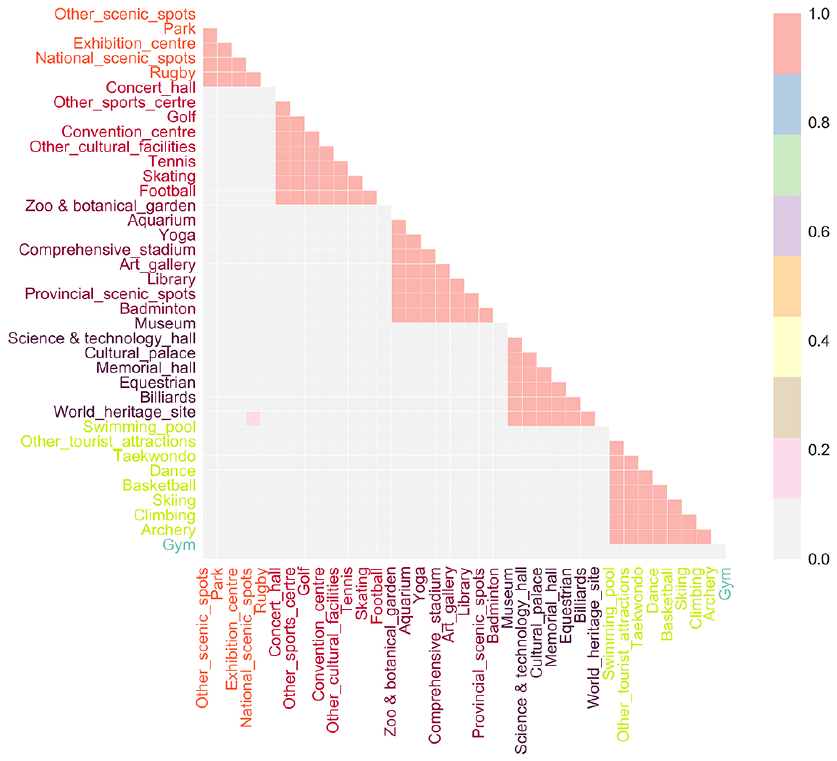}
	\caption{Cosine Similarity between Venue Categories.}
\end{figure}

\begin{figure}
	\includegraphics{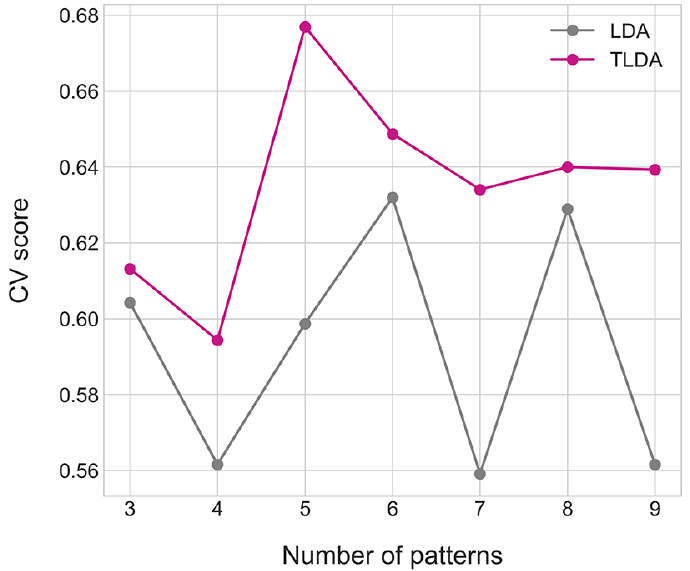}
	\caption{Comparison between TLDA and LDA (Venues).}
\end{figure}

To further evaluate the clustering performance of the model, we compare the CV scores \cite{roder2015exploring} of TLDA with that of LDA. Since the LDA model does not contain the temporal part extended by TLDA, only the venue category clustering is evaluated. Again, we run the analysis iteratively with $K$ setting as 3-9 and present the results in Figure 7. We can see that the TLDA model outperforms LDA in all the seven cases. This result indicates that the TLDA model not only enriches LDA by considering temporal features, it is also superior to classical LDA by generating more coherent topics. In addition to venue category information, TLDA also provides us another point of view to learn about cultural patterns temporally. The temporal characteristics of the cultural patterns are displayed and compared in Figure 8. For comparison, values are shown in percentage to present the degrees to which time periods are representative for the patterns. To present the hourly patterns more clearly, we group 24 hours into five slots, which are morning (6-11am), noon (11am-14pm), afternoon (14-19pm), evening (19-24pm), and night (0-6am).

\begin{figure*}
	\includegraphics{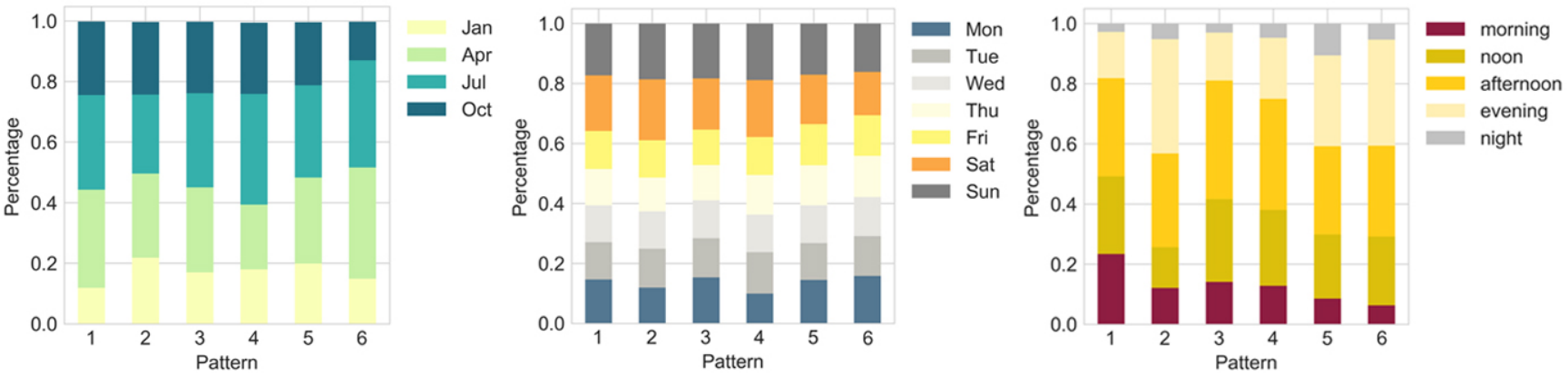}
	\caption{Temporal Characteristics of Cultural Patterns in Beijing.}
\end{figure*}

The results in Figure 5 and Figure 8 together reveal the key characteristics of the six cultural patterns detected in Beijing. We can observe that pattern one is composed of people who love travelling and prefer wide open space. They do not like visiting scenic spots and parks during winter very much, perhaps due to harsher weather conditions. Compared to other groups, the first pattern has the highest percentage of activity in the morning and the lowest during nighttime. This finding can be linked with what we can expect from real life experience, that parks in Beijing are always full of people doing morning exercises. Music fans make up the majority of pattern 2. For this group of people, their cultural visiting frequencies during the four seasons are rather balanced. However, on weekends and evening, they present considerably higher probability of being active when compared to their counterparts. This can be explained by the fact that concerts are usually being scheduled and attended in the evening hours. The third pattern are nature lovers who like plants and animals in particular. This group of people present hourly sensitive features as they prefer visiting cultural places during the daytime to evening or at night. Then, the forth pattern corresponds to museum lovers. This cultural group are particularly active in summer and in the afternoons. Moreover, they have the lowest percentage of activity on Monday compared to other patterns. This phenomenon can probably be explained by the fact that many museums are closed on Mondays. The fifth group are sports fans, especially swimming enthusiasts. These people have the highest percentage of nighttime activity. The last group of people are gym lovers with a single top cultural venue category becoming prominent here with an extremely high percentage of 0.99. Spring and summer time is the most popular period for them to exercise. They do not like going to the gym in the morning, preferring evenings in most cases. Additionally, even though weekends make the greatest contributions to almost all the cultural patterns, they are not prominent in the case of pattern 6.

\subsection{Refined Cultural Planning for Beijing}

After uncovering the cultural patterns, we map how the demand and supply levels of each pattern are distributed spatially, and calculate the demand-supply ratio for urban areas. In this part of analysis, we divide the city into 400m by 400m grids aligned with the latitude and longitude dimensions. Each cell is called an area, and the centroid of which is used to represent the cell's demand-supply balance. 

From the demand respective, we begin with the application of the POPTICS algorithm to find centres of activity range for six groups of cultural fans in their daily lives based on all the check-ins they created previously.  Then, we collect a particular subgroup of cultural check-ins for each cultural fan according to the pattern he belongs to. We find his influential radius $ r_{u_z} $, and calculate the demand value he contributes to his surrounding areas based on equation (13). After the need of all users in a certain pattern group are aggregated by formula (14), the overall demand for each cultural pattern can be obtained as the first row in Figure 9 suggests. With respect to the supply side, we calculate the supply capability of each area in terms of various cultural patterns according to equations (15) and (16). The supply levels categorised by patterns across the city are visualised in the middle row subfigures, followed by the final demand-supply ratio output shown in the last row in Figure 9. For the first two rows, the darker the colour, the higher the demand or supply level; while for the DSR, red and blue represent high and low ratios, respectively. As can be observed from the figure, the demand for different cultural patterns is distributed in a similar manner spatially. The highest demand areas cluster in the urban area between the 2nd and 5th ring road, while some hotspots shown in suburbs areas like Yanqing, Huairou, and Miyun Districts. When we look at the supply level, six patterns present a more heterogeneous behaviour. Although a general pattern can be discovered that the cultural supply capabilities show a decreasing trend from the city centre to the suburbs, this inequality is less obvious in patterns 1, 4 and 5. From our final results of DSR, we can find the inner city inside the former city walls (Xicheng and Dongcheng) is in great need of cultural services of pattern 1 and 5, like parks, swimming pools, and exhibition centres. For pattern 2 and 3, the demand-supply gaps of related cultural facilities are relatively equal within the city, while for the last pattern, the need for gym service is relatively greater in outer suburbs.

Through the demand-supply analysis above using the DSI model, we get priority lists of urban areas in terms of different types of cultural services according the levels of need. To validate our model, we calculate the Pearson correlation between the DSR value and the average distance users need to travel for a particular kind of cultural services. The correlation coefficients for six patterns are presented in Figure 10. From this figure, we can see that five patterns show high positive correlations except pattern one. The distinctive result observed in pattern one can probably be explained by its top venue categories. As an ancient city, many of the scenic spots in Beijing are historical relics, the locations of which are not decided by modern urban planning. The overall result of the correlation analysis suggests that users from areas in great needs of a type of cultural services generally have to travel longer distances to be served. It further indicates that the facilities in the users' surrounding areas are not enough to fulfill their needs, and thus provides evidence for our results that services in high priority areas detected by the models indeed are insufficient compared to their counterparts. 

\begin{figure*}
	\includegraphics{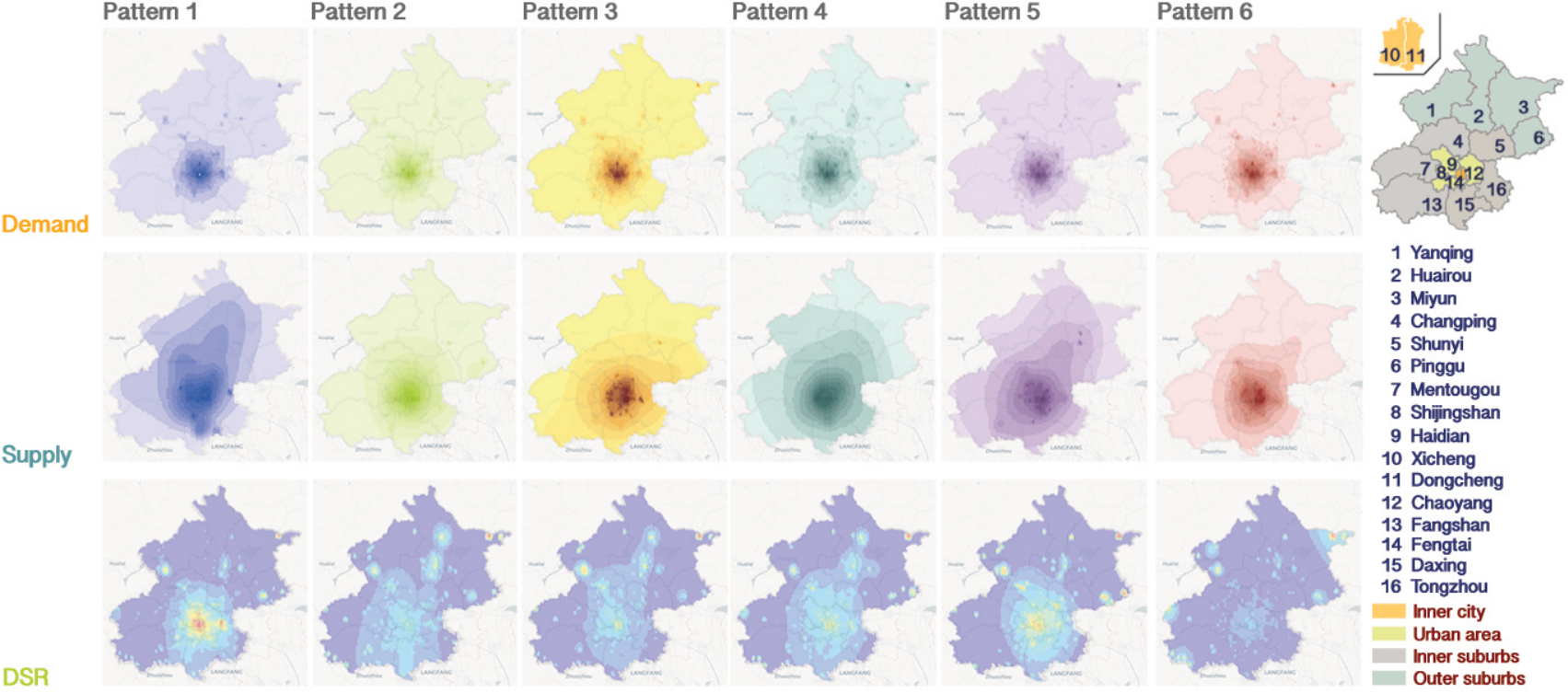}
	\caption{Results of DSI Model.}
\end{figure*}

\begin{figure}
	\includegraphics{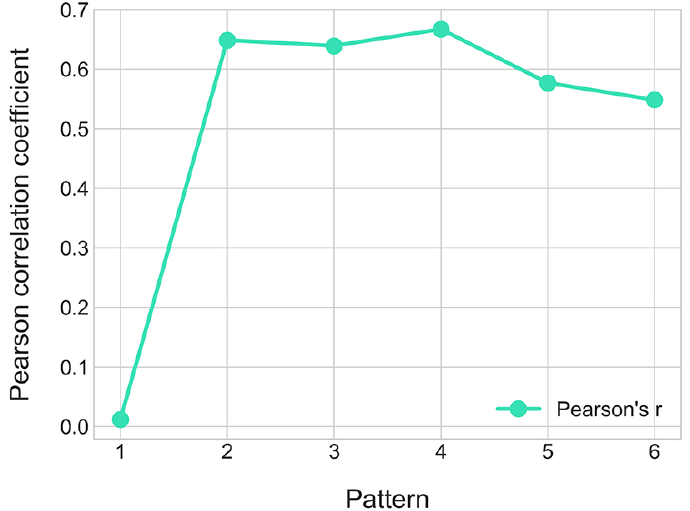}
	\caption{Correlation between DSR and Travel Distance.}
\end{figure}

\section{Conclusion}
In this paper we have proposed a data-driven framework for urban cultural planning. The framework exploits a time-aware topic model to identify latent patterns of urban cultural interactions. Using then a density-based algorithm named POPTICS, we identify the primary locations of activity of mobile users and couple this with the TLDA output to generate cartographic representations indicative of the demand-supply balance for cultural resources in the city. We evaluate our approach using implicit user feedback, demonstrating how user active in areas that lack cultural establishments bear larger transportation costs to access cultural resources. Besides urban policy makers, the findings of this resesarch can also provide suggestions to business owners on the opening hours, and to citizens on neighbourhood characteristics in the city. Overall, we demonstrate how the new generation of datasets emerging through modern location-based systems can provide an edge in city planning as they offer rich views on urban mobility dynamics and allow for the development of population adaptive frameworks that move beyond static representations of area-level population densities. 

\section{Acknowledgements}
	We would like to thank Tencent for hosting Xiao Zhou and providing her with access to the datasets for this study. The first author acknowledges the financial support co-funded by the China Scholarship Council and the Cambridge Trust.

\bibliographystyle{plain}
{\small

}


\end{document}